\documentclass[twocolumn,showpacs,aps,prb,superscriptaddress,floatfix]{revtex4}
\usepackage{graphicx}
\usepackage{dcolumn}
\usepackage{bm}

\begin{document}

\preprint{}




\author{K.C. Rule}
\affiliation{Department of Physics and Astronomy, McMaster University,
Hamilton, Ontario, L8S 4M1, Canada}
\author{M.J. Lewis}
\affiliation{Department of Physics and Astronomy, McMaster University,
Hamilton, Ontario, L8S 4M1, Canada}
\author{H.A. Dabkowska}
\affiliation{Department of Physics and Astronomy, McMaster University,
Hamilton, Ontario, L8S 4M1, Canada}
\author{D.R. Taylor}
\affiliation{Department of Physics, Engineering Physics, and Astronomy, Queen's University, Kingston, Ontario, K7L 3N6, Canada}
\author{B.D. Gaulin}
\affiliation{Department of Physics and Astronomy, McMaster University,
Hamilton, Ontario, L8S 4M1, Canada}
\affiliation{Canadian Institute for Advanced Research, 180 Dundas St. W.,
Toronto, Ontario, M5G 1Z8, Canada}

\newcommand{\LBCO}{La$_{2-x}$Ba$_x$CuO$_{4}$ }
\newcommand{\LSCO}{La$_{2-x}$Sr$_x$CuO$_{4}$ }
\newcommand{\TBTO}{Tb$_2$Ti$_2$O$_7$ }
\title{Critical X-ray Scattering Studies of Jahn-Teller Phase Transitions in TbV$_{1-x}$As$_{x}$O$_{4}$.}
\date{\today}

\begin{abstract}
The critical behaviour associated with cooperative Jahn-Teller phase transitions in TbV$_{1-x}$As$_{x}$O$_{4}$ (where \textit{x} = 0, 0.17, 1) single crystals have been studied using high resolution x-ray scattering.  These materials undergo continuous tetragonal $\to$ orthorhombic structural phase transitions driven by Jahn-Teller physics at T$_C$ = 33.26(2) K, 30.32(2) K and 27.30(2) K for \textit{x} = 0, 0.17 and 1 respectively.  The orthorhombic strain was measured close to the phase transition and is shown to display mean field behavior in all three samples.  Pronounced fluctuation effects are manifest in the longitudinal width of the Bragg scattering, which diverges as a power law, with an exponent given by $x=0.45 \pm 0.04$, on approaching the transition from either above or below.  All samples exhibited twinning; however the disordered x = 0.17 sample showed a broad distribution of twins which were stable to relatively low temperatures, well below T$_C$.  This indicates that while the orthorhombic strain continues to develop in a conventional mean field manner in the presence of disorder, twin domains are easily pinned by the quenched impurities and their associated random strains.  \\
\end{abstract}


\maketitle 

\section{Introduction}
TbVO$_{4}$ and TbAsO$_{4}$ belong to a group of compounds which undergo a structural phase transition induced by the cooperative Jahn-Teller effect. Both TbVO$_{4}$ and TbAsO$_{4}$ exhibit a transition from the tetragonal \textit{I}4$_{1}$/\textit{amd} structure at high temperatures to the twinned orthorhombic \textit{Fddd} phase at temperatures below 35 K\cite{hutchings}. At room temperature the low-lying crystal electric field levels for the Tb$^{3+}$ ions in the TbVO$_4$ environment consist of two singlets separated by 2.23 meV and a degenerate, non-Kramers doublet midway between the singlets\cite{elliot,Gehring75}. The Jahn-Teller phase transition changes the local Tb$^{3+}$ environment, with a concomitant splitting of the doublet,  and increase of the singlet separation to 6.32 meV\cite{Gehring75}. The resulting magnetic ground state for the Tb$^{3+}$ ions can then order magnetically at much lower temperatures\cite{schafer}. The structural phase transitions have also been studied in disordered systems wherein the non-magnetic cations are mixed. Such disordered materials as TbV$_{1-x}$As$_x$O$_4$ are characterized by the presence of quenched random strains introduced by the ionic radius mismatch between V$^{5+}$ and As$^{5+}$, which is quite large $\sim$ 20$\%$.

These materials and their Jahn-Teller phase transitions have been known and studied for some time.  For example, optical birefrigence studies of the critical properties of TbVO$_4$ and DyVO$_4$ showed that the former displayed mean field criticality near its Jahn-Teller transition, while the latter displayed 3D Ising-like universality near its transition\cite{harley}.  Early x-ray experiments measured the order parameter relevant to these Jahn-Teller transitions, the orthorhombic strain, but were not able to give accurate critical property measurements for this family of terbium oxides\cite{will,smith}.  An x-ray scattering study of the related TbV$_{1-x}$P$_x$O$_4$ system\cite{hirano} has been carried out, although again, the emphasis was not on the critical regime.

The role of quenched disorder on the Jahn-Teller phase transitions in the TbV$_{1-x}$As$_x$O$_4$ system has more recently been the subject of further optical birefrigence studies, which also show changes in criticality induced by the presence of impurities\cite{choo}.  The change in criticality has been discussed in the context of the Random Field Ising Model\cite{imry} and how it may be realized in a structural analogue of the more-familiar case arising in magnetism - that of the dilute Ising antiferromagnet in the presence of a uniform magnetic field\cite{belanger}.

We have carried out a high resolution x-ray scattering study of the tetragonal to orthorhombic Jahn-Teller phase transitions in the TbV$_{(1-x)}$As$_x$O$_4$ system, with particular emphasis on the critical regime, where the orthorhombic strain is small and fluctuations in the strain may be expected to be large.  This will enable a quantitative microscopic understanding of the critical phenomena in the pure, end members of the family, and show the extent to which the critical behaviour in the mixed system is modified by the presence of quenched disorder.

\section{Experimental Details}
Single crystals of TbVO$_{4}$ and TbAsO$_{4}$ and their solid solutions were grown from PbO/PbF$_{2}$ flux by the slow cooling method. Pre-annealed starting materials Tb$_{2}$O$_{3}$ and X$_{2}$O$_{5}$ where X is either V or As were weighed and combined with PbO and PbF$_{2}$, which act as high temperature solvents. The mix was then melted in a Pt crucible at 1300$^\circ$C and slow cooled (1-2 deg/h) to 900$^\circ$C \cite{jasiolek,hagenmuller}.  Single crystals in the form of elongated rods with typical dimensions of 5x2x2 mm were separated from this melt. The TbV$_{0.83}$As$_{0.17}$O$_4$ single crystal studied was the same one previously studied by Schriemer et al \cite{schriemer}.  In that study the relative As concentration in TbV$_{1-x}$As$_x$O$_4$ is given as x=0.15, in good agreement with the value x=0.17 which we have determined from the current x-ray measurements.

X-ray scattering measurements were performed using a rotating-anode, Cu K$_\alpha$ x-ray source and a four circle diffractometer employing a two dimensional area detector.  The single crystals were mounted within a helium-filled sample cell which was attached to the cold finger of a closed-cycle helium refrigerator.  Scattering measurements were carried out as a function of temperature with a temperature stability of $\sim$ 0.005 K, appropriate for critical scattering studies.  CuK$_{\alpha1}$ radiation from the 18 kW rotating anode x-ray generator was selected using a perfect single crystal Ge (110) monochromator.  The diffracted beam was measured using a Bruker HiStar area detector, mounted on the scattering arm at a distance of 0.7 m from the sample.  This configuration allowed for high resolution characterization of the Bragg peaks.  The scattered intensity was measured as a function of temperature for both the (6,6,0) Bragg reflection in the tetragonal phase, which splits into the (12,0,0) and (0, 12, 0) reflections in the twinned orthorhombic phase, as well as for the (8,0,0) Bragg reflection in the tetragonal phase, which splits into (8, 8, 0) and (8, -8, 0) reflections in the same twinned orthorhombic phase.


\begin{figure}
\centering
\includegraphics[width=1\linewidth]{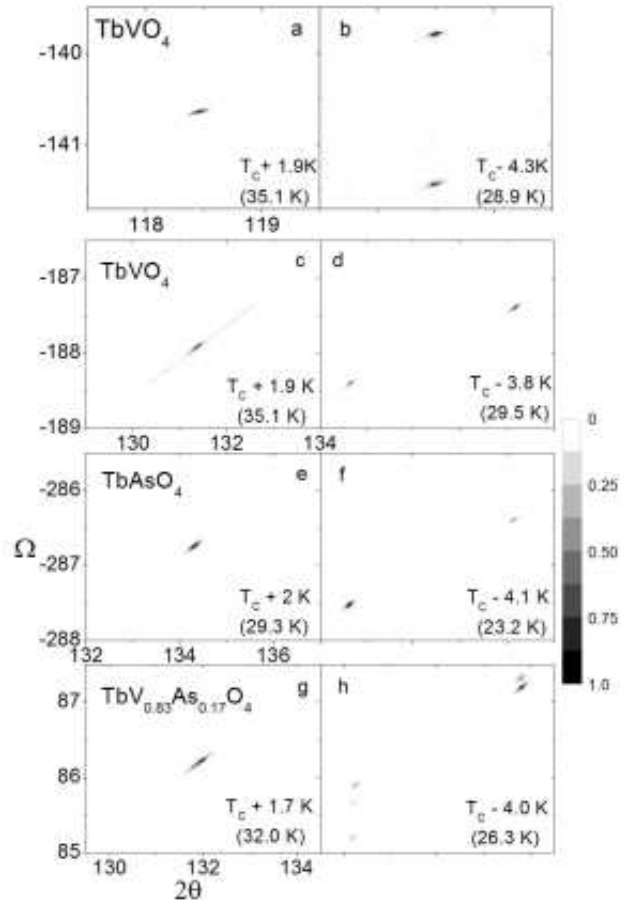}
\caption{Representative contour maps of the temperature dependence of the Bragg peaks which split on going through the Jahn-Teller phase transitions in TbVO$_{4}$, TbAsO$_{4}$ and TbV$_{0.83}$As$_{0.17}$O$_{4}$.  (a) and (b) show the splitting of the (8, 0, 0) Bragg peak into (8, 8, 0)/(8, -8, 0) Bragg peaks in TbVO$_4$. (c) and (d), (e) and (f), and (g) and (h) show the splitting of the 
(6, 6, 0) Bragg peak into the (12, 0, 0) and (0, 0, 12) Bragg peaks in TbVO$_{4}$, TbAsO$_{4}$ and TbV$_{0.83}$As$_{0.17}$O$_{4}$, respectively.  The data on the left ((a), (c), (e) and (g)) are above T$_{C}$ while the data on the right ((b), (d), (f) and (h)) are below T$_{C}$.  The diagonal line in (b) indicates the cut used for the longitudinal scans in Figure 5.  A vertical line would indicate a transverse scan.}   
\label{Figure 1}
\end{figure}

\section{Measurements of the Orthorhombic Strain}

Figure 1 shows representative x-ray scattering contour maps taken for TbVO$_{4}$, TbAsO$_{4}$ and TbV$_{0.83}$As$_{0.17}$O$_{4}$ above and below their appropriate phase transition temperatures.  Figure 1(a) and (b) shows data taken around the (8, 8, 0) Bragg position (in the orthorhombic phase) in TbVO$_{4}$, while Figures 1 (c) and (d), (e), and (f), and (g) and (h) shows data taken around the (12, 0, 0) Bragg positions (in the orthorhombic phase) in TbVO$_{4}$, TbAsO$_{4}$, and TbV$_{0.83}$As$_{0.17}$O$_{4}$, respectively.  All of these scattering data sets were acquired by integrating the Bragg scattering in the vertical direction for each sample rotation angle ($\Omega$, plotted on the y-axis of Fig. 1), and then plotting the resulting intensity as a function of scattering angle, 2$\theta$.  The dashed line in Fig. 1(c)  indicates a longitudinal scan, while a vertical line on any of these panels would indicate a transverse scan.  The transverse splitting of the (8, 0, 0) tetragonal Bragg peak into (8, 8, 0) orthorhombic Bragg peaks in TbVO$_{4}$ is shown in Fig. 1(a) and (b), while the longitudinal splitting of the tetragonal (6, 6, 0) Bragg peaks into twin orthogonal peaks at (12, 0, 0) and (0, 12, 0) in all three Tb-based oxides is seen in Figs 1(c) - (h).  

These data sets can be analysed by extracting the positions for all the Bragg peaks in the field of view in both scattering angle, 2$\theta$, and sample orientation angle, $\Omega$.  The scattering angle for the single crystal peaks in both the tetragonal phase above T$_C$, and the twinned orthorhombic phase below T$_C$, is used to extract the relevant lattice parameters and orthorhombic strain, defined as the difference between the a and b lattice parameters.  The scattering angles for all of the (6, 6, 0) (tetragonal phase)  and (12, 0, 0) (orthorhombic phase) Bragg peaks of all three samples are shown as a function of temperature in Fig. 2.   Figure 3 (a) shows the correponding orthorhombic strain below T$_C$ in all three samples.  Figure 3 (b) shows the splitting in domain orientation for TbVO$_{4}$, TbAsO$_{4}$ and TbV$_{0.83}$As$_{0.17}$O$_{4}$, extracted from the difference in sample orientation, $\Omega$ positions, of the Bragg peaks from the principal (12, 0, 0) and (0, 12, 0) orthorhombic twin domains below T$_C$.

Figure 2 (a)  shows the splitting of (12, 0, 0) and (0, 12, 0) Bragg peaks in all three systems in absolute units of scattering angle, while Fig. 2 (b) shows the same data, but now relative to the scattering angle of the Bragg peak in the high temperature tetragonal (HTT) phase.  This data shows quite clearly that all three samples display continuous phase transitions, as was known previously for TbVO$_{4}$\cite{will} and TbAsO$_{4}$\cite{berkhan}, and that their critical properties are similar.  The scattering in Fig. 2 (a) however, also shows that within the high temperature tetragonal phase, the lattice parameter (identified by its scattering angle)  for  TbV$_{0.83}$As$_{0.17}$O$_{4}$ lies $\sim$ 17$\%$ of the way between TbVO$_{4}$ and TbAsO$_{4}$, as given by a linear variation in lattice parameter between the end members of the series, TbVO$_{4}$ and TbAsO$_{4}$.  Such a linear variation is observed in the TbV$_{1-x}$P$_x$O$_4$ system\cite{hirano}.  This establishes an accurate concentration for the TbV$_{0.83}$As$_{0.17}$O$_{4}$ single crystal by microscopic means.

Figure 3 (a) shows the corresponding orthorhombic strains, defined as the difference in lattice parameters a and b, as a function of temperature below T$_C$ for each of TbVO$_{4}$, TbAsO$_{4}$ and TbV$_{0.83}$As$_{0.17}$O$_{4}$.  Related measurements of the orthorhombic strain have been measured previously as a function of temperature for the TbV$_{1-x}$P$_x$O$_4$ system\cite{hirano}.  However, the present measurements are taken sufficiently close to the phase transitions such that an accurate determination of T$_C$ and a quantitative study of the critical properties of these phase transitions is possible.  It is clear from the data that the order parameter, the orthorhombic strain, rises very sharply with decreasing temperature, such that little of this data would reside in the asymptotic critical regime, where the order parameter is small compared with its saturation value.  Therefore we fit this data to a full mean field solution, Eq. 1, rather than simply to the asymptotic expression for small values of the order parameter.  This gives a mean field critical exponent for the order parameter, $\beta$ =0.5, but the full data sets are accurately described.  This description of the order parameters as a function of temperature using the solution to the transcendental Eq. 1 below, is shown as the solid lines in Fig. 3 (a) and (b).   Clearly the full mean field solution provides an excellent description of this data. 

\begin{equation}
\label{1}
m=tanh[{{qJm}\over{k_BT}}]
\end{equation}

Figure 3 (b) shows the evolution of the splitting between the twin domain orientation angles, $\Omega$, describing the (12, 0, 0) and (0, 12, 0) twin orthorhombic domains below T$_C$ in TbVO$_{4}$, TbAsO$_{4}$ and TbV$_{0.83}$As$_{0.17}$O$_{4}$.  Again the temperature dependence of these splittings is very well described by the full mean field solution to the order parameter, Eq. 1.  This plot follows the evolution of the domain orientation for both the majority and minority domains present within the field-of-view of our scattering experiments, for which typical data sets are shown in Fig. 1.  For TbVO$_{4}$ and TbAsO$_{4}$, a single majority and minority domain is visible at most temperatures. However the TbV$_{0.83}$As$_{0.17}$O$_{4}$ data sets are qualitatively different with many domains visible even at temperatures well removed from T$_C$.  For example, the data set for TbV$_{0.83}$As$_{0.17}$O$_{4}$ at T=26.3 K (T$_C$-4 K) shows a total of 5 twin domains; 2 majority twin domains at high scattering angle and 3 minority twin domains at low scattering angle.  Moreover the dynamic range in domain orientation angle spanned by the multiple twin domains is considerably larger than is found in either TbVO$_{4}$ or TbAsO$_{4}$ below T$_C$.  Figure 3 (b) quantifies this behaviour, following the splitting of the majority and minority twin domain orientation angles ($\Omega$) below T$_C$.  The extrapolated, low temperature $\Omega$ splitting is very similar for TbVO$_{4}$ and TbAsO$_{4}$, but considerably larger for TbV$_{0.83}$As$_{0.17}$O$_{4}$, where the splitting between the orientation angle, $\Omega$, of the most intense minority twin and most intense majority twin domain is plotted.

The order parameter data shown in Fig. 3, along with the fits to full mean field behaviour, Eq. 1, allow a precise determination of the critical temperature T$_C$, in these systems.  Fitting both the orthorhombic strain shown in Fig. 3 (a) and the twin domain orientation splitting shown in Fig. 3 (b), we obtain T$_C$ = 33.26(2) K, 30.32(2) K and 27.30(2) K  for TbVO$_{4}$, TbV$_{0.83}$As$_{0.17}$O$_{4}$ and TbAsO$_{4}$, respectively.  In itself, this is a somewhat surprising result, as it shows the T$_C$ for TbV$_{0.83}$As$_{0.17}$O$_{4}$ to lie midway between that of TbVO$_{4}$ and  TbAsO$_{4}$, despite only 17$\%$ of the V ions being replaced by As ions.  This is very likely a manifestation of the quenched impurities on the Jahn-Teller driven phase transition in TbV$_{0.83}$As$_{0.17}$O$_{4}$.

The behaviour of the twin domains below T$_C$ in TbV$_{0.83}$As$_{0.17}$O$_{4}$ can be qualitatively understood in terms of pinning of the twin domains by random strains due to the quenched impurities.  Clearly, this does not influence the critical properties of the order parameter, the orthorhombic strain, which remains well described by mean field behaviour at all temperatures, as is the case for both TbVO$_{4}$ and TbAsO$_{4}$.  The influence of quenched impurities on twin domain interfaces has been investigated by light scattering techniques\cite{schriemer}, which is expected to be sensitive to such large scale structure.  The present results are qualitatively consistent with these results, showing stable multi-domain twin states in TbV$_{0.83}$As$_{0.17}$O$_{4}$, distinct from TbVO$_{4}$ and  TbAsO$_{4}$, and therefore a larger volume fraction taken up by domain interfaces in TbV$_{0.83}$As$_{0.17}$O$_{4}$.  However, the change in critical properties previously reported\cite{choo} on introduction of quenched impurities is not observed.




\begin{figure}
\centering
\includegraphics[width=1\linewidth]{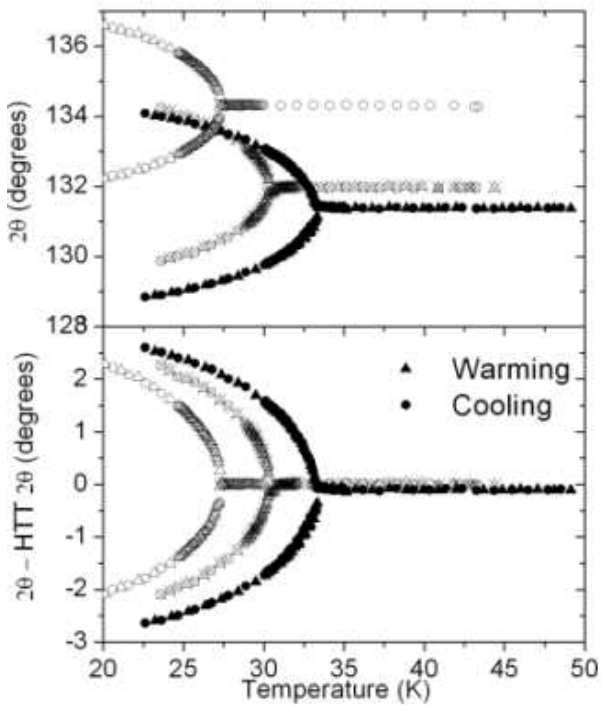}
\caption{The temperature dependence of the peak position in $2\theta$ for each of the three samples: TbVO$_{4}$ (closed symbols), TbAsO$_{4}$ (open symbols) and TbV$_{0.83}$As$_{0.17}$O$_{4}$ (crossed symbols).  The structural phase transition is indicated by the splitting of the single Bragg intensity into two with decreasing temperature. Figure 2a shows the true $2\theta$ splitting for each of the compounds, while Figure 2b shows the splitting relative to the scattering in the tetragonal phase at high temperatues.}   
\label{Figure 2}
\end{figure}


\begin{figure}
\centering
\includegraphics[width=1\linewidth]{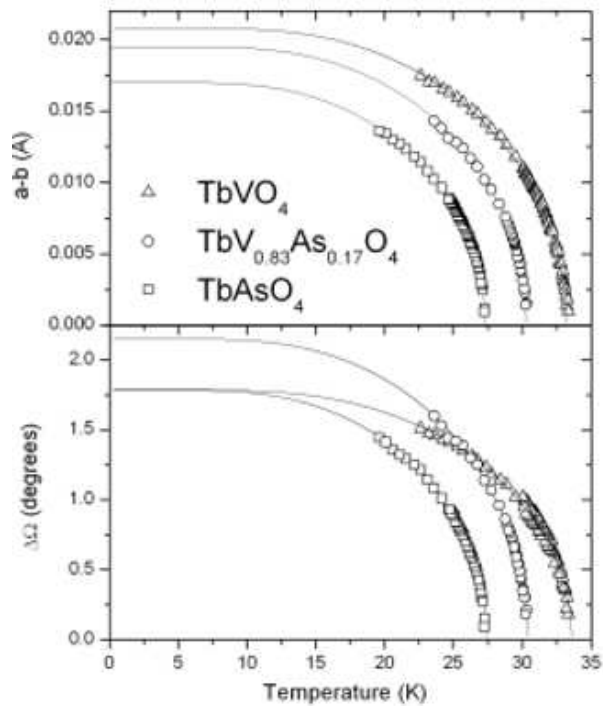}
\caption{Figure 3a (top) shows the measured order parameters, the orthorhombic strains (a-b), in TbVO$_4$,  TbV$_{0.83}$As$_{0.17}$O$_{4}$, and TbAsO$_4$.  These have been fit to the mean field solution (Equation 1) for the order parameter as a function of temperature, which is shown as the solid lines.  Figure 3b (lower) shows the domain orientation $\Omega$ splitting of the (12, 0, 0) and (0, 12, 0) Bragg peaks and fits to the same mean field solution.  The TbV$_{0.83}$As$_{0.17}$O$_{4}$ $\Omega$ splitting was taken as the difference between the majority and minority domains with the strongest Bragg intensity.  This was an issue for TbV$_{0.83}$As$_{0.17}$O$_{4}$ alone as the twinned domains collapsed into a single dominant majority and minority domain peak within a few degrees of T$_C$ for the pure materials.}  
\label{Figure 3} 
\end{figure}

\begin{figure}
\centering
\includegraphics[width=1\linewidth]{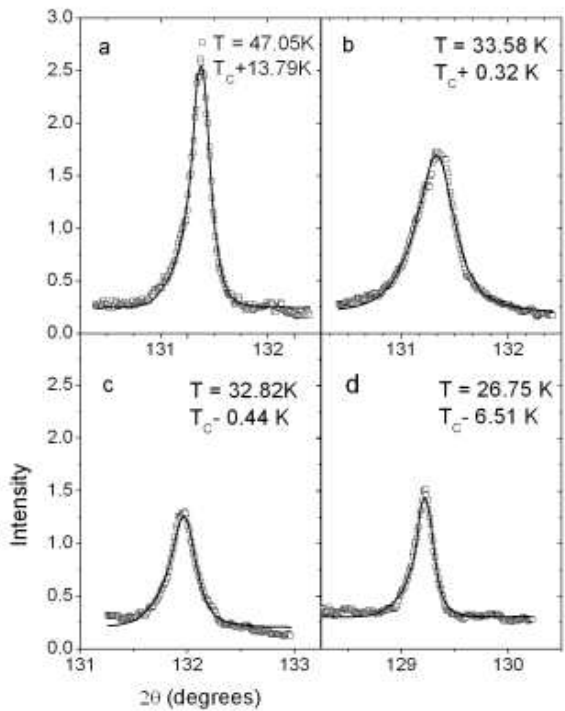}
\caption{Representative fits to the longitudinal scans for TbVO$_{4}$.  This data is taken from longitudinal cuts to the full data sets, as shown in Fig. 1 (c).  (a) and (d) were taken well above and below T$_C$=33.26(2) K, respectively, and are resolution limited.  (b) and (c) are relatively close to T$_C$ and show substantial broadening in the longitudinal direction. The resolution-corrected longitudinal widths 
of the profiles were extracted and are plotted as a function of temperature in Figs. 5 and 6.}   
\label{Figure 4}
\end{figure}

\section{Measurements of Fluctuations of the Orthorhombic Strain}

Close examination of the scattering data sets at the (12, 0, 0)/(0, 12, 0) Bragg positions in both TbVO$_{4}$ and TbAsO$_{4}$ revealed a very interesting longitudinal broadening in the vicinity of T$_C$.  Representative longitudinal scans near the (6, 6, 0) Bragg position in the tetragonal phase and (12, 0, 0) Bragg positions in the orthorhombic phase of TbVO$_{4}$ are shown in Fig. 4.  Very similar data was obtained for TbAsO$_{4}$, but it is not displayed.  Figure 4 (a) shows a longitudinal scan through (6, 6, 0) well above T$_C$ at T=47.05 K (T$_C$+ 13.8 K), while Fig. 4 (d) shows a longitudinal scan through (12, 0, 0) at T=26.75 K, 6.51 K below T$_C$.  Both of these scans have approximately the same intrinsic width, and we consider them to be resolution limited.  Figures 4 (b) and (c) however, show longitudinal scans much closer to T$_C$. Figure 4 (b) shows data just above T$_C$ at T=33.58 K (T$_C$+ 0.32 K) while Fig. 4 (c) shows the same scan just below T$_C$, at T=32.82 K (T$_C$ - 0.44 K).  It is clear that both of these longitudinal scans near T$_C$ are appreciably broader than those well removed from T$_C$, and therefore are not resolution limited.

We associate this broadening with fluctuations in the order parameter, similar to that observed near continuous magnetic phase transitions.  However, in the present case, fluctuations in the orthorhombic strain lead to a longitudinal broadening of Bragg peaks of the form (6, 6, 0) in the tetragonal phase.  Rather than narrowing in angle or reciprocal space as the phase transition is approached, these peaks broaden and decrease in peak intensity as T$_C$ is approached from above.  Broadening is also observed below T$_C$, but the widths of the Bragg peaks fall off quickly, such that all the scattering in both TbVO$_{4}$ and TbAsO$_{4}$ is resolution limited for temperatures more than 2 K below T$_C$.

Data of the form shown in Fig. 4 was fit to an Ornstein-Zernike form, Eq. 2, convoluted with the resolution function appropriate to the measurement.  We employed the scattering well above T$_C$, at T $\sim$ 50 K, as the resolution function, assuming that all fluctuation effects are gone at such high temperatures.  The results of fitting the TbVO$_{4}$ data is shown as the solid lines in Fig. 4, and this clearly provides an excellent description of the longitudinal scans at all temperatures,  Similar quality fits to the TbAsO$_{4}$ were also obtained.  The resulting intrinsic widths to the longitudinal scans are shown 
as a function of temperature for TbVO$_{4}$ in Fig. 5 (a) and for TbAsO$_{4}$ in Fig. 5 (b).  Data is shown for both independent warming and cooling runs for both materials, although no significant history dependence to the widths of the scattering were observed.  For temperatures below T$_C$, longitudinal scans through both majority and minority twin domains were investigated.  Over the narrow range of temperature below T$_C$ for which a finite width can be observed, the intrinsic widths of the majority and minority twin domain peaks differ, with the minority twin peak showing a greater intrinsic width than the majority twin domain, consistent with the interpretation of the minority twin domains as being relatively small.  A related set of data near T$_C$  was acquired for TbV$_{0.83}$As$_{0.17}$O$_{4}$, however the relevant longitudinal lineshapes were more complicated than those observed in TbVO$_{4}$ and TbAsO$_{4}$, and these results are not discussed here.

\begin{equation}
S({2\theta})=
\frac{A}{(2\theta-2\theta_{peak})^2 +\kappa ^2} 
\end{equation}

The most striking and interesting result is that the intrinsic width of the Bragg peaks, interpreted as an inverse correlation length, displays a pronounced peak at T$_C$, reminiscent of $\lambda$  anomolies in the heat capacity of systems undergoing continuous phase transitions.  The temperature dependence of the divergence of the intrinsic width, or inverse correlation length, is examined on a log-log plot in Fig. 6, wherein data above T$_C$ in TbVO$_{4}$ and TbAsO$_{4}$ is plotted vs reduced temperature (T-T$_C$)/T$_C$.  Again independent warming and cooling runs are shown, with no history dependence observed.  This plot, which employs the same T$_C$s used to describe the critical properties of the order parameter shown in Fig. 3, clearly brings out the power law nature of the divergence of the intrinsic width of the scattering as T$_C$ is approached from above.  Figure 6 (a) shows this data for both TbVO$_{4}$ and TbAsO$_{4}$, and it is clear that both materials display a very similar power law divergence  over at least one and a half decades in reduced temperature.  Figure 6 (b) shows the intrinsic width of the longitudinal scan in TbAsO$_{4}$ alone, where a single power law describes its divergence over almost three decades in reduced temperature.  The exponent characterizing the divergence is x=0.45 $\pm$ 0.04.

While we have characterized the power law divergence of the inverse correlation length, or intrinsic width, at T$_C$, we do not have a precise interpretation for this critical phenomenon.  In their high-resolution gamma-ray diffraction experiments on TbVO$_{4}$, Smith and Tanner\cite{smith} observed a broadening in the rocking-curve widths, that is the transverse width of the Bragg peaks, as the phase transition was approached from above that somewhat resembles the broadening that we measured. They did not analyze it in detail, but suggested that it arose from the strains associated with random internal stresses relaxing as the elastic constant C$_{66}$ tends to zero. While such a mechanism could contribute in our case, it is difficult to believe that it would exhibit critical behaviour over such a wide range and with such well defined power law behaviour as we observe. 

As mentioned previously, our observed longitudinal broadening of the (6, 6, 0) (tetragonal) and the (12, 0, 0)/(0, 12, 0) (orthorhombic) Bragg peaks in Fig. 5 greatly resembles heat capacity anomolies associated with continuous phase transitions.  This may be a natural explanation of the intrinsic width, as the width is inversely proportional to the fluctuating domain size, and thus proportional to the volume fraction of the material occupied at any one time by interface.  The energy of such a system would scale as the amount of interface, and a longitudinal Bragg peak width scaling as a derivative of the energy, the heat capacity, seems physically resonable.  The critical exponent x may be close to that describing fluctuations near a mean field, or mean field tricritical point, for which a large heat capacity exponent characteristic of a divergence is appropriate.  At present we cannot be more precise, but hope that the observation and characteriozation of this critical behaviour informs and motivates further work and understanding.


\begin{figure}
\centering
\includegraphics[width=1\linewidth]{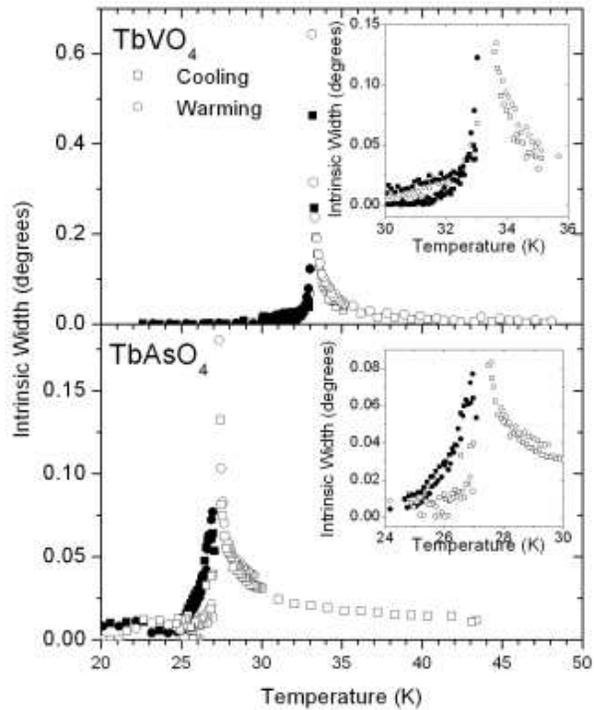}
\caption{Resolution-corrected longitudinal widths, or inverse correlation lengths, showing critical fluctuations around the (6, 6, 0) (tetragonal phase) and (12, 0, 0) and (0, 12, 0) (orthorhombic phase) of TbVO$_{4}$ (top) and TbAsO$_{4}$ (bottom).  Below T$_{C}$ open data points refer to the majority twin domain, while closed data points refer to the minority twin domain.  Square and circle data points 
refer to independent cooling and warming cycles.}   
\label{Figure 5}
\end{figure}

\begin{figure}
\centering
\includegraphics[width=1\linewidth]{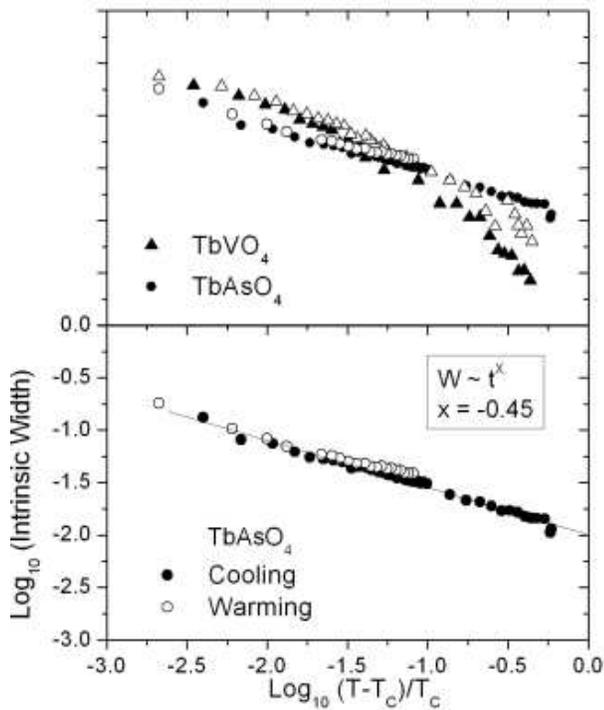}
\caption{(top) Critical behaviour of the longitudinal width, $\kappa$ from Eq. 2, above T$_C$ for TbVO$_4$ and TbAsO$_4$ are shown as a function of reduced temperature.  The same data
for TbAsO$_4$, along with a fit to a power law divergence characterized by the exponent x=-0.45 $\pm$ 0.04.}   
\label{Figure 6}
\end{figure}


  
\section{Comparison to Related Structural Phase Transitions and Fluctuations}

We can compare the present results on the order parameter and fluctuations near the Jahn-Teller tetragonal to orthorhombic structural phase transitions in TbVO$_{4}$, TbAsO$_{4}$ and TbV$_{0.83}$As$_{0.17}$O$_{4}$, to those in related systems.  The principal characteristics of the present results are mean field critical behaviour for the orthorhombic strain, and the observation of
clear fluctuation effects in the vicinity of T$_C$.

Two topical and well studied systems which undergo tetragonal to orthorhombic structural phase transitions are the high temperature superconductors \LBCO and \LSCO.  Critical x-ray and neutron scattering on these materials do not show mean field behaviour, but rather the orthorhombic strain goes like the square of the order parameter\cite{lsco,zhao}, which is well described by three dimensional universality.  Theoretical expectations for these systems are for 3D XY critical behaviour\cite{3dxy}, and this is consistent with what is observed\cite{lsco,zhao}.  Even qualitatively, it is clear that the growth in the orthorhombic strain with decreasing temperature in TbVO$_{4}$, TbAsO$_{4}$ and TbV$_{0.83}$As$_{0.17}$O$_{4}$ is different from that observed in \LBCO and \LSCO.  Comparing to a recent high resolution x-ray scattering study on the \LBCO system\cite{zhao}, we can see that the orthorhombic strain grows in a relatively gentle manner with decreasing temperature compared with what is observed in TbVO$_{4}$, TbAsO$_{4}$ and TbV$_{0.83}$As$_{0.17}$O$_{4}$.  Figure 3 shows that the latter orthorhombic strains grow to $\sim$ half of their saturation values at 0.9 $\times$ T$_C$.  The same growth in the \LBCO system requires lower temperatures and half of saturation is only achived by $\sim$ 0.7 T$_C$.

There are differences of course, between the high temperature superconducting copper oxides and the terbium based oxides we report here.  The cuprates are quasi-two-dimensional, for example, while TbVO$_{4}$, TbAsO$_{4}$ and TbV$_{0.83}$As$_{0.17}$O$_{4}$ are three dimensional.  Most notably however the magnetic Cu$^{2+}$ ion in \LBCO and \LSCO is not Jahn-Teller active - a single hole resides in a non-degenerate d orbital.  This would appear to be the most likely source for the difference in critical behaviour.  TbVO$_{4}$, TbAsO$_{4}$ and TbV$_{0.83}$As$_{0.17}$O$_{4}$  possess Tb$^{3+}$ ions whose orbitals and the occupation of the orbitals change on going through T$_C$, likely generating long range strain fields with concomittant mean field behaviour.

Another interesting comparison is to other Tb$^{3+}$-based transition metal oxides.  Here an interesting comparison can be made with a very recent high resolution x-ray scattering study of the cubic pyrochlore \TBTO\cite{ruff}.  This geometrically-frustrated magnetic material does not undergo a structural phase transition at finite temperature, but recent x-ray scattering measurements observe broadening of the allowed structural Bragg peaks at low temperatures.  These have been interpreted as fluctuations above a very low temperature Jahn-Teller phase transition, which is either never realized, or only realized at unattainably low temperatures.  Indeed, to our knowledge, the present work is the only other study of structural phase transitions which show such critical broadening of the Bragg peaks in the vicinity of T$_C$.  The temperature scale over which these fluctuations are observed is quite different in \TBTO as compared with TbVO$_{4}$, TbAsO$_{4}$ and TbV$_{0.83}$As$_{0.17}$O$_{4}$.  In the present case, the broadening of the Bragg peaks appears as critical phenomena and the strongest fluctuation effects fall off within a few degrees of T$_C$.  In contrast, the Jahn-Teller-like fluctuations in \TBTO occur coincident with the formation of a spin liquid state, and are measureable over an extended regime of temperature from 20 K to 0.3 K.

Finally we return to the issue of mean field criticality vs criticality associated with random fields.  We did not observe a departure from mean field criticality associated with random fields in TbV$_{0.83}$As$_{0.17}$O$_{4}$.   This may be because the random fields associated with the disorder are too weak to influence the critical behaviour in the range of reduced temperatures that we
could access.  A measure of the strength of the disorder and associated random fields is the extent to which they depress the transition temperature. The observed T$_C$ for our mixed sample was 
30.32 K, lower than the value 32.25 K expected from a linear variation in T$_C$ between the two end members. This depression in T$_C$ confirms that disorder is relevant and that associated random
fields are present. In some related systems that have been studied, DyAs$_{x}$V$_{1-x}$O$_{4}$\cite{graham,reza} and KH$_{2}$As$_{x}$P$_{1-x}$O$_{4}$\cite{taylor05}, however, as well as dilute antiferromagnetic systems\cite{belanger} the depressions in T$_C$ were larger and random-field critical behaviour was clearly observed.
 
\section{conclusions}

We have studied the critical properties near Jahn-Teller, tetragonal to orthorhombic structural phase transitions in TbVO$_{4}$, TbAsO$_{4}$ and TbV$_{0.83}$As$_{0.17}$O$_{4}$ using high resolution x-ray scattering.  These results show the order parameter, the orthorhombic strain, in all three materials to exhibit mean field criticality.  The disordered member of this series, TbV$_{0.83}$As$_{0.17}$O$_{4}$ is distinguished only in so far as it exhibits a wide dynamic range of majority and minority orthorhombic twin domain orientations, and these are stable to low temperatures.  We ascribe this to the pinning of the domain structure by random strains arising from the quenched disorder in TbV$_{0.83}$As$_{0.17}$O$_{4}$.

We also observe very interesting fluctuation effects in TbVO$_{4}$ and TbAsO$_{4}$ in the vicinity of T$_C$.  These are manifest in longitudinal broadening of the (6, 6, 0) Bragg peaks in the tetragonal phase, which split to become (12, 0, 0) and (0, 12, 0) Bragg peaks in the orthorhombic low temperature phase.  This broadening exhibits a power law divergence at T$_C$, in TbVO$_{4}$ and TbAsO$_{4}$, with a critical exponent, x=0.45 $\pm$ 0.04.  We hope that this latter work informs and motivates a full theoertical understanding of this fluctuation phenomena.

This work was supported by NSERC of Canada.

\end{document}